\documentclass[12pt,fleqn]{book} % Default font size and left-justified equations
\usepackage[top=1.0in,bottom=1.0in,left=1.0in,right=1.0in,headsep=10pt,letterpaper]{geometry} % Page margins
\usepackage{xcolor} % Required for specifying colors by name
\definecolor{ocre}{RGB}{52,177,201} % Define the orange color used for highlighting throughout the book

\usepackage{avant} % Use the Avantgarde font for headings
\usepackage{mathptmx} % Use the Adobe Times Roman as the default text font together with math symbols from the Sym­bol, Chancery and Computer Modern fonts

\usepackage{microtype} % Slightly tweak font spacing for aesthetics
\usepackage[utf8]{inputenc} % Required for including letters with accents
\usepackage[T1]{fontenc} % Use 8-bit encoding that has 256 glyphs

\usepackage{titlesec} % Allows customization of titles

\usepackage{graphicx} % Required for including pictures
\graphicspath{{Pictures/}{figures/cosmo/}} % Specifies the directory where pictures are stored

\usepackage{lipsum} % Inserts dummy text

\usepackage{tikz} % Required for drawing custom shapes

\usepackage[english]{babel} % English language/hyphenation

\usepackage{enumitem} % Customize lists
\setlist{nolistsep} % Reduce spacing between bullet points and numbered lists

\usepackage{booktabs} % Required for nicer horizontal rules in tables

\usepackage{eso-pic} % Required for specifying an image background in the title page

\usepackage{titletoc} % Required for manipulating the table of contents

\contentsmargin{0cm} % Removes the default margin
\titlecontents{chapter}[1.25cm] % Indentation
{\addvspace{15pt}\large\sffamily\bfseries} % Spacing and font options for chapters
{\color{ocre!60}\contentslabel[\Large\thecontentslabel]{1.25cm}\color{ocre}} % Chapter number
{}
{\color{ocre!60}\normalsize\sffamily\bfseries\;\titlerule*[.5pc]{.}\;\thecontentspage} % Page number
\titlecontents{section}[1.25cm] % Indentation
{\addvspace{5pt}\sffamily\bfseries} % Spacing and font options for sections
{\contentslabel[\thecontentslabel]{1.25cm}} % Section number
{}
{\sffamily\hfill\color{black}\thecontentspage} % Page number
[]
\titlecontents{subsection}[1.25cm] % Indentation
{\addvspace{1pt}\sffamily\small} % Spacing and font options for subsections
{\contentslabel[\thecontentslabel]{1.25cm}} % Subsection number
{}
{\sffamily\;\titlerule*[.5pc]{.}\;\thecontentspage} % Page number
[]

\titlecontents{lsection}[0em] % Indendating
{\footnotesize\sffamily} % Font settings
{}
{}
{}

\titlecontents{lsubsection}[.5em] % Indentation
{\normalfont\footnotesize\sffamily} % Font settings
{}
{}
{}

\usepackage{fancyhdr} % Required for header and footer configuration

\fancypagestyle{plain}{\fancyhead{}} % Style for when a plain pagestyle is specified

\makeatletter
\renewcommand{\cleardoublepage}{
\clearpage\ifodd\c@page\else
\hbox{}
\vspace*{\fill}
\thispagestyle{empty}
\newpage
\fi}

\usepackage{amsmath,amsfonts,amssymb,amsthm} % For math equations, theorems, symbols, etc

\newtheoremstyle{ocrenumbox}% % Theorem style name
{0pt}% Space above
{0pt}% Space below
{\normalfont}% % Body font
{}% Indent amount
{\small\bf\sffamily\color{ocre}}% % Theorem head font
{\;}% Punctuation after theorem head
{0.25em}% Space after theorem head
{\small\sffamily\color{ocre}\thmname{#1}\nobreakspace\thmnumber{\@ifnotempty{#1}{}\@upn{#2}}% Theorem text (e.g. Theorem 2.1)
\thmnote{\nobreakspace\the\thm@notefont\sffamily\bfseries\color{black}---\nobreakspace#3.}} % Optional theorem note
\newtheoremstyle{blacknumex}% Theorem style name
{5pt}% Space above
{5pt}% Space below
{\normalfont}% Body font
{} % Indent amount
{\small\bf\sffamily}% Theorem head font
{\;}% Punctuation after theorem head
{0.25em}% Space after theorem head
{\small\sffamily{\tiny\ensuremath{\blacksquare}}\nobreakspace\thmname{#1}\nobreakspace\thmnumber{\@ifnotempty{#1}{}\@upn{#2}}% Theorem text (e.g. Theorem 2.1)
\thmnote{\nobreakspace\the\thm@notefont\sffamily\bfseries---\nobreakspace#3.}}% Optional theorem note

\newtheoremstyle{blacknumbox} % Theorem style name
{0pt}% Space above
{0pt}% Space below
{\normalfont}% Body font
{}% Indent amount
{\small\bf\sffamily}% Theorem head font
{\;}% Punctuation after theorem head
{0.25em}% Space after theorem head
{\small\sffamily\thmname{#1}\nobreakspace\thmnumber{\@ifnotempty{#1}{}\@upn{#2}}% Theorem text (e.g. Theorem 2.1)
\thmnote{\nobreakspace\the\thm@notefont\sffamily\bfseries---\nobreakspace#3.}}% Optional theorem note

\newtheoremstyle{ocrenum}% % Theorem style name
{5pt}% Space above
{5pt}% Space below
{\normalfont}% % Body font
{}% Indent amount
{\small\bf\sffamily\color{ocre}}% % Theorem head font
{\;}% Punctuation after theorem head
{0.25em}% Space after theorem head
{\small\sffamily\color{ocre}\thmname{#1}\nobreakspace\thmnumber{\@ifnotempty{#1}{}\@upn{#2}}% Theorem text (e.g. Theorem 2.1)
\thmnote{\nobreakspace\the\thm@notefont\sffamily\bfseries\color{black}---\nobreakspace#3.}} % Optional theorem note
\makeatother

\newcounter{dummy}
\numberwithin{dummy}{section}
\theoremstyle{ocrenumbox}
\newtheorem{theoremeT}[dummy]{Theorem}

\newtheorem{exerciseT}{Exercise}[chapter]
\theoremstyle{blacknumex}
\newtheorem{exampleT}{Example}[chapter]
\theoremstyle{blacknumbox}

\newtheorem{definitionT}{Definition}[section]
\newtheorem{corollaryT}[dummy]{Corollary}
\theoremstyle{ocrenum}

\RequirePackage[framemethod=default]{mdframed} % Required for creating the theorem, definition, exercise and corollary boxes

\newmdenv[skipabove=7pt,
skipbelow=7pt,
backgroundcolor=black!5,
linecolor=ocre,
innerleftmargin=5pt,
innerrightmargin=5pt,
innertopmargin=5pt,
leftmargin=0cm,
rightmargin=0cm,
innerbottommargin=5pt]{tBox}

\newmdenv[skipabove=7pt,
skipbelow=7pt,
rightline=false,
leftline=true,
topline=false,
bottomline=false,
backgroundcolor=ocre!10,
linecolor=ocre,
innerleftmargin=5pt,
innerrightmargin=5pt,
innertopmargin=5pt,
innerbottommargin=5pt,
leftmargin=0cm,
rightmargin=0cm,
linewidth=4pt]{eBox}	

\newmdenv[skipabove=7pt,
skipbelow=7pt,
rightline=false,
leftline=true,
topline=false,
bottomline=false,
linecolor=ocre,
innerleftmargin=5pt,
innerrightmargin=5pt,
innertopmargin=0pt,
leftmargin=0cm,
rightmargin=0cm,
linewidth=4pt,
innerbottommargin=0pt]{dBox}	

\newmdenv[skipabove=7pt,
skipbelow=7pt,
rightline=false,
leftline=true,
topline=false,
bottomline=false,
linecolor=gray,
backgroundcolor=black!5,
innerleftmargin=5pt,
innerrightmargin=5pt,
innertopmargin=5pt,
leftmargin=0cm,
rightmargin=0cm,
linewidth=4pt,
innerbottommargin=5pt]{cBox}

\makeatletter
\renewcommand{\@seccntformat}[1]{\llap{\textcolor{ocre}{\csname the#1\endcsname}\hspace{1em}}}
\renewcommand{\section}{\@startsection{section}{1}{\z@}
{-2ex \@plus -1ex \@minus -.2ex}
{1ex \@plus.1ex }
{\normalfont\large\sffamily\bfseries}}
\renewcommand{\subsection}{\@startsection {subsection}{2}{\z@}
{-2ex \@plus -0.1ex \@minus -.2ex}
{0.5ex \@plus.2ex }
{\normalfont\sffamily\bfseries}}
\renewcommand{\subsubsection}{\@startsection {subsubsection}{3}{\z@}
{-2ex \@plus -0.1ex \@minus -.2ex}
{.2ex \@plus.2ex }
{\normalfont\small\sffamily\bfseries}}
\renewcommand\paragraph{\@startsection{paragraph}{4}{\z@}
{-2ex \@plus-.2ex \@minus .2ex}
{.1ex}
{\normalfont\small\sffamily\bfseries}}

\usepackage{hyperref}
\hypersetup{hidelinks,backref=true,pagebackref=true,hyperindex=true,colorlinks=false,breaklinks=true,urlcolor= ocre,bookmarks=true,bookmarksopen=false,pdftitle={Title},pdfauthor={Author}}

\newcommand{\thechapterimage}{}
\newcommand{\chapterimage}[1]{\renewcommand{\thechapterimage}{#1}}

\def\thechapter{\arabic{chapter}}
\def\@makechapterhead#1{
\thispagestyle{empty}
{\centering \normalfont\sffamily
\ifnum \c@secnumdepth >\m@ne
\if@mainmatter
\startcontents
\begin{tikzpicture}[remember picture,overlay]
\node at (current page.north west)
{\begin{tikzpicture}[remember picture,overlay]
\node[anchor=north west,inner sep=0pt] at (0,0) {\includegraphics[width=\paperwidth]{\thechapterimage}};
\draw[anchor=west] (5cm,-9cm) node [rounded corners=20pt,fill=ocre!10!white,text opacity=1,draw=ocre,draw opacity=1,line width=1.5pt,fill opacity=.6,inner sep=12pt]{\huge\sffamily\bfseries\textcolor{black}{\thechapter. #1\strut\makebox[22cm]{}}};
\end{tikzpicture}};
\end{tikzpicture}}
\par\vspace*{230\p@}
\fi
\fi}

\def\@makeschapterhead#1{
\thispagestyle{empty}
{\centering \normalfont\sffamily
\ifnum \c@secnumdepth >\m@ne
\if@mainmatter
\begin{tikzpicture}[remember picture,overlay]
\node at (current page.north west)
{\begin{tikzpicture}[remember picture,overlay]
\node[anchor=north west,inner sep=0pt] at (0,0) {\includegraphics[width=\paperwidth]{\thechapterimage}};
\draw[anchor=west] (5cm,-6cm) node [rounded corners=20pt,fill=ocre!10!white,fill opacity=.6,inner sep=12pt,text opacity=1,draw=ocre,draw opacity=1,line width=1.5pt]{\LARGE\sffamily\bfseries\textcolor{black}{#1\strut\makebox[22cm]{}}};
\end{tikzpicture}};
\end{tikzpicture}}
\par\vspace*{130\p@}
\fi
\fi
}
\makeatother

\usepackage{graphicx}
\graphicspath{{./figures/}}
\usepackage{appendix}
\usepackage{latexsym,amsmath,amsfonts,amssymb,booktabs}
\usepackage[font=small]{caption}
\usepackage{slashed,upgreek,amscd,cancel,tensor,color}
\usepackage{adjustbox}
\usepackage[numbers,sort&compress,square]{natbib}
\usepackage{epsfig,latexsym}
\usepackage{url}
\numberwithin{equation}{section}% numera le equazioni seconde le sezioni , e.g. 1.15 invece che consecutivamente; anche le appendici, eq. (A.1) etc. Richiede amsmath
\usepackage{doi}
\usepackage{subcaption}
\usepackage{mathtools}
\usepackage{upgreek}
\usepackage{stfloats}
\usepackage{afterpage}
\usepackage{multirow}

\usepackage{aas_macros}
\usepackage{tcolorbox}
\usepackage[utf8]{inputenc}

\begin{document}

\chapterimage{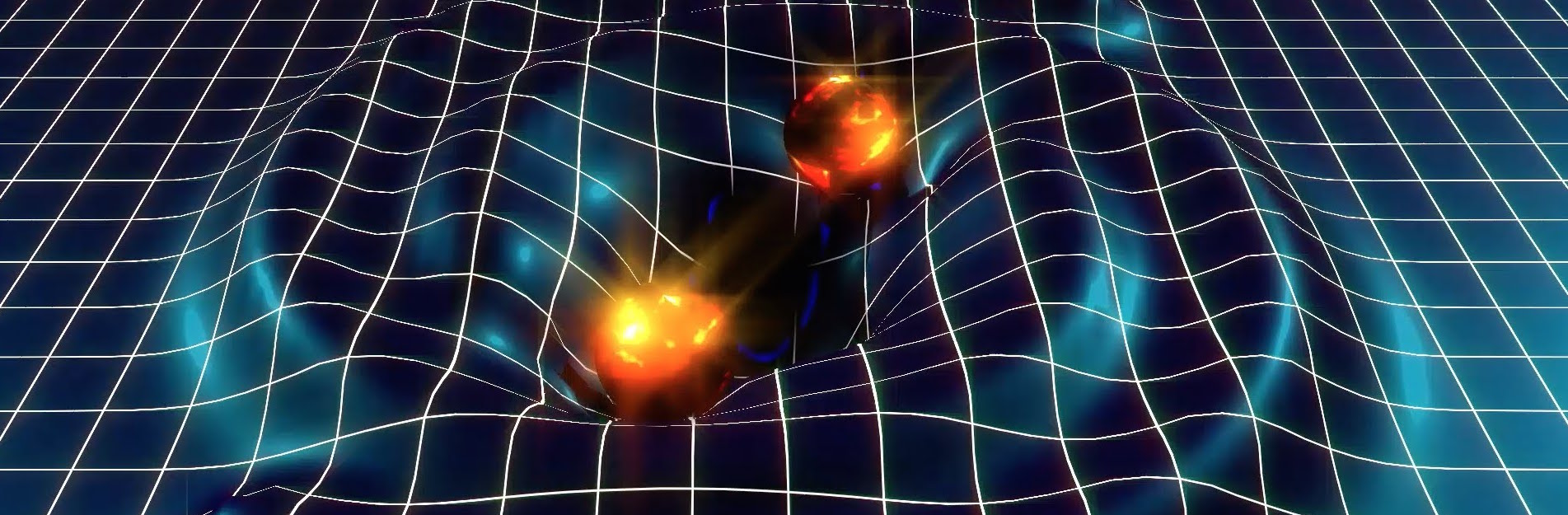} % Chapter heading image
\chapter*{\Large Binary Black Holes: Deeper, Wider, Sharper} 

%\begin{figure*}\center
%  \includegraphics[width=1.0\textwidth]{warped-spacetime.jpg}
%\end{figure*} 
  
%\raggedright
\begin{center} 
\Large
\textbf{Astro2020 Science White Paper} \linebreak

%\vspace{-2cm} 

DEEPER, WIDER, SHARPER: \linebreak NEXT-GENERATION GROUND-BASED GRAVITATIONAL-WAVE OBSERVATIONS \linebreak OF BINARY BLACK HOLES \linebreak

\end{center} 

\normalsize

\vspace{-3cm}

\noindent \textbf{Thematic Areas:} 
\begin{itemize}
\item Formation and Evolution of Compact Objects 
\item Stars and Stellar Evolution 
\item Multi-Messenger Astronomy and Astrophysics
\end{itemize}

% Planetary Systems 
% Star and Planet Formation 
% Formation and Evolution of Compact Objects 
% Cosmology and Fundamental Physics 
% Stars and Stellar Evolution 
% Resolved Stellar Populations and their Environments 
% Galaxy Evolution   
% Multi-Messenger Astronomy and Astrophysics 

\vspace{0.75cm}

\noindent \textbf{Principal Author}: 

\vspace{-0.3cm}
 Name:	Vicky Kalogera
 \newline
 	
	\vspace{-0.8cm}					
Institution: Northwestern U.\ 
 \newline
 
 \vspace{-0.8cm}
Email: vicky@northwestern.edu
 \newline
 
 \vspace{-0.8cm}
Phone: +1-847-491-5669 
 \newline
 
 \noindent 
\textbf{Lead Co-authors:} Christopher P.L.\ Berry (Northwestern U.), Monica Colpi (U.\ of Milano -- Bicocca), Steve Fairhurst (Cardiff U.), Stephen Justham (UCAS, Beijing and U.\ of Amsterdam), Ilya Mandel (Monash U.), Alberto Mangiagli (U.\ of Milano -- Bicocca), Michela Mapelli (U.\ of Padova), Cameron Mills (Cardiff U.), B.S. Sathyaprakash (Penn State U.\ and Cardiff U.), Raffaella Schneider (INFN Roma), Thomas Tauris (Aarhus U.), Rosa Valiante (INAF Roma)

 \noindent 
Click here for \href{https://docs.google.com/spreadsheets/d/1uNKEW77Fm-_nc21_3jSOX-P4ecRxsxA5UgDOD4bkG9Y/edit#gid=0}{\bf other co-authors and supporters}

%(names and institutions)

 % \linebreak

%\textbf{Abstract :}

%ADD ABSTRACT
\pagebreak

%-------- coverpage, copyright and table of contents

%\chapterimage{intro.jpg} % Chapter heading image
\chapterimage{warped-spacetime.jpg} % Chapter heading image
%\chapter*{\LARGE Formation and Evolution of Compact Objects}

%\chapter*{Deeper, wider, sharper: Next-generation GW observations of binary black holes}

\section*{Gravitational-wave discovery space for black holes}% at high frequencies

The Advanced LIGO gravitational-wave (GW) detector network began observations in 2015 \cite{Abbott:2016blz}. 
Since then, the first two observing runs (later including Advanced Virgo) 
have yielded the discovery of ten binary black hole (BBH) systems and one binary neutron star system \cite{LIGOScientific:2018mvr}. 
Already these detections have revolutionized astrophysics of stellar-mass black holes (BHs) \cite{TheLIGOScientific:2016htt,LIGOScientific:2018jsj}, provided new tests of general relativity \cite{TheLIGOScientific:2016src,Yunes:2016jcc,Abbott:2017vtc,Abbott:2017oio,Abbott:2018lct}, and launched the field of multimessenger GW astronomy \cite{Monitor:2017mdv,Abbott:2017wuw}. 

Through to the end of the next decade, this detector network will continue to be enhanced as sensitivities reach design goals and new detectors come online \cite{Aasi:2013wya}. 
In the BBH domain, we will be able to detect a pair of $10 M_\odot$ BHs out to a redshift of $z \simeq 1$ \cite{Aasi:2013wya}. 
The annual BBH detection rates are forecast to be several hundreds of mergers and science benefits will compound through accumulated observing time and growing detected samples \cite{Mandel:2009nx,Trifiro:2015zda,Stevenson:2017dlk,Talbot:2017yur,Barrett:2017fcw,Zevin:2017evb}.

Beyond this horizon, step-wise sensitivity improvements with the \emph{next generation of ground-based GW observatories} will be required if we are to pursue major science questions that cannot be answered by the current and near-term GW facilities \cite[e.g.,][]{Sathyaprakash:2012jk,Evans:2016mbw}. 
Current-generation GW detectors are able to provide constraints on the merger-rate densities in the local Universe and approximate distributions of component masses \cite{LIGOScientific:2018jsj}; 
however, precise measurements of, for example, spin magnitudes and tilts are of paramount importance to understand their origin and the evolutionary physics of the binary system \cite{Mandel:2009nx,Rodriguez:2016vmx,Vitale:2015tea,Stevenson:2017dlk,Talbot:2017yur,Qin:2018nuz}. 
This information is essential to obtain insights on the formation channels of compact binaries.
While instrumental designs are an active area of research, we highlight here \emph{how next-generation GW ground-based detectors will enable us to survey deeper, to observe a wider range of frequencies, and to make more precise physical measurements and will transform the study of BBH astrophysics}.

\begin{tcolorbox}[standard jigsaw,colframe=ocre,colback=ocre!10!white,opacityback=0.6,coltext=black]
Next-generation GW observations will uncover BBHs throughout the entire Universe back to the beginning of star formation, and will detect new source types (if they exist) beyond stellar-mass binaries, such as intermediate-mass black holes. 
\begin{itemize}[leftmargin=*]
\item {\bf Discover binary black holes throughout the observable Universe.} 
What is the merger rate as a function of redshift to the beginning of the reionization era, and how does it correlate with massive star formation, metallicity, and galaxy evolution? 

\item {\bf Reveal the fundamental properties of black holes.} 
What are the mass and spin demographics of black holes throughout the Universe, are they correlated, and do they evolve with redshift? What do they reveal about the formation and evolutionary origin of BBHs? 

\item {\bf Uncover the seeds of supermassive black holes.} 
Do intermediate-mass black hole mergers occur in nature, and can such black holes serve as the long sought seeds of supermassive black holes? Is there a single thread which connects the formation of stellar-mass and supermassive black holes?
\end{itemize}
\end{tcolorbox}

\section*{Deeper -- A survey of black holes throughout cosmic time}

With a next-generation GW detector network, for the \emph{first time}, we will detect BH mergers at redshifts beyond $z \sim 1$ and we will measure the evolution of the BBH merger rate out to redshifts of $z \gtrsim 10$ \cite{Aasi:2013wya,Fishbach:2018edt,Vitale:2018yhm}: \emph{over the entire history of the Universe}. 
GW astronomy would thereby gain a synoptic view of the evolution of BHs across cosmic time, beyond the peak in star-formation rate at $z \sim 2$ \cite{Madau:2014bja} back to the cosmic dawn around $z \sim 20$ when the Universe was only $200~\mathrm{Myr}$ old.

Measurements of merger rate vs.\ redshift combined with measurements of the BHs' physical properties at unprecedented accuracies will enable conclusive constraints on BBH formation channels. 
Stellar-origin BBH formation tracks cosmic star formation \cite{Dominik:2013tma,Mapelli:2017hqk,Chruslinska:2018hrb,Mapelli:2018wys}, while the density of primordial BHs is not expected to correlate with the star formation density \cite{Sasaki:2018dmp,Scelfo:2018sny};
different binary channels are predicted to lead to different distributions of delay times between formation and merger \cite{Dominik:2012kk,Kinugawa:2015nla,Mandel:2015qlu,Marchant:2016wow,Rodriguez:2016avt,Barrett:2017fcw,Fragione:2018vty,Rodriguez:2018rmd,Choksi:2018jnq}. 
Therefore, determining the merger rate as a function of redshift provides a unique insight into the lives of BBHs.
\emph{Only next-generation GW detectors can survey the complete redshift range of merging BBHs and provide a sufficiently large catalog of detections to constrain the full BBH population and their origins}.

\begin{figure}[!h]
\center
  \includegraphics[width=0.49\textwidth]{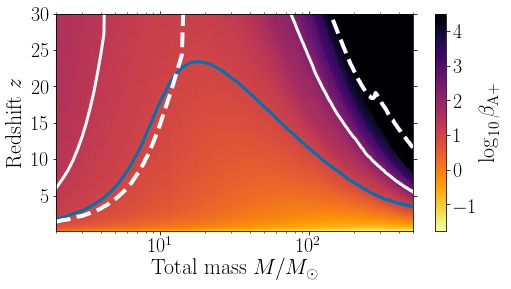}
  \includegraphics[width=0.49\textwidth]{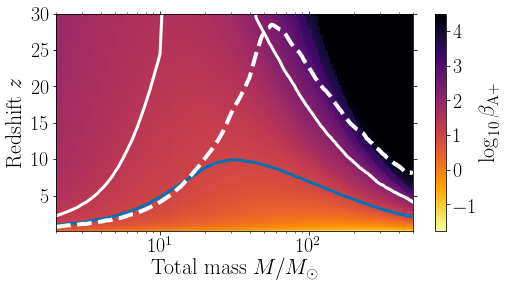}
  \caption{Color maps show the boost factor relative to the LIGO A+ design $\beta_\mathrm{A+}$ required to see a binary with a given total source mass $M$ out to given redshift. 
The color bar saturates at $\log_{10} \beta_\mathrm{A+} = 4.5$; some high-mass systems at high redshift are not detectable for any boost factor as there is no signal above $5~\mathrm{Hz}$. 
Panels are for mass ratios $q = 1$ (\textbf{left}) and $q = 0.1$ (\textbf{right}). 
The blue curve highlights the reach at a boost factor of $\beta_\mathrm{A+} = 10$. 
The solid and dashed white lines indicate the maximum reach of Cosmic Explorer \cite{Evans:2016mbw} and the Einstein Telescope \cite{Sathyaprakash:2012jk}, respectively; sources below these curves would be detectable.}
\label{BCO-detect}
\end{figure}

To capture BBH mergers across the stellar mass spectrum (up to total masses of $M \simeq 200 M_\odot$) all the way back to the end of the cosmological dark ages ($z \simeq 20$), a major advance in GW detector sensitivity is required.
This cannot be delivered by the maximal sensitivity planned for the current ground-based detector facilities. 
We quantify this sensitivity step by the \emph{boost factor} $\beta_\mathrm{A+}$ relative to the LIGO A+ design \cite{Barsotti:2018} between $5~\mathrm{Hz}$ and $5~\mathrm{kHz}$ (and no sensitivity outside this range).
In Figure~\ref{BCO-detect}, we show this boost factor, required to detect an optimally-oriented, overhead binary at a signal-to-noise ratio (SNR) of $8$, as a function of the binary's total mass and redshift.

The boost factors $\beta_\mathrm{A+}$ needed to acquire a complete census of BBH mergers throughout the Universe are well within the design aspirations for next-generation designs such as Cosmic Explorer \cite{Evans:2016mbw} and the Einstein Telescope \cite{Sathyaprakash:2012jk}; 
for these specific sensitivity assumptions, BBH mergers of total mass $M \sim 10$--$40 M_\odot$ can be detected out to $z \sim 10^2$.

Observations of the cosmological distribution of coalescing binaries would complement planned electromagnetic surveys designed to study stars and stellar remnants back to cosmic dawn \cite{Whalen:2012yk,Koopmans:2015sua,Cassano:2018zwm,Kalirai:2018qfg,Katz:2019akl}, as well as millihertz GW observations made by \textit{LISA} \cite{Audley:2017drz}, which can observe systems ranging from local stellar-mass binaries (days to years before they enter the frequency range of terrestrial detectors) \cite{Sesana:2016ljz,AmaroSeoane:2009ui} to supermassive black hole (SMBH) systems in the centers of galaxies \cite{Klein:2015hvg,Babak:2017tow}.
\textit{Athena} \cite{Barret:2013bna} and the mission concept \textit{Lynx} \cite{LynxTeam:2018usc} would detect SMBHs back to high redshift ($z \gtrsim 7$);   
\textit{Lynx} would observe $10^3 M_\odot$ BHs to $z \sim 5$ and $10^2 M_\odot$ BHs to $z \sim 2$, while \textit{Athena} would survey these in the nearby Universe.  
Next-generation GW detectors have the unique potential to observe stellar-mass BH systems back to the early Universe.

\section*{Wider -- Expanding the black hole mass spectrum}

Electromagnetic astronomy has benefited enormously from advancing observing facilities to cover an expanded range of frequencies. 
These enable new probes of previous known sources, and allow for the discovery of new types of previously unobserved sources. 
\emph{Next-generation GW detectors have the unique capability to push the frequency range down to $\simeq 1~\mathrm{Hz}$ and up to $\simeq 5~\mathrm{kHz}$}, while improving performance across the band in between. 

The merger frequency for a coalescing binary scales inversely with the mass of the binary, 
hence observing at lower frequency opens up the potential of detecting more massive BHs. 
Reaching down to frequencies of $\simeq 1~\mathrm{Hz}$ is the most robust means to prove the existence of intermediate-mass black holes (IMBHs) in binaries with masses in excess of $100 M_\odot$ \cite{Huerta:2010tp,Haster:2015cnn}.
The discovery of IMBHs \cite{Gair:2010dx,Abbott:2017iws} would be uniquely impactful: these could be formed through dynamical processes in star clusters \cite{2015MNRAS.454.3150G,Mapelli:2016vca} or from the collapse of massive metal-poor stars \cite{Madau:2001sc,Ball:2011mp,Ryu:2016dou}, 
and may potentially be the seed BHs which grow into SMBHs \cite{Volonteri:2012tp,Latif:2016qau,Bernal:2017nec,Woods:2018lty}. 
SMBHs are observed up to redshift $z=7.54$ \cite{Banados:2017unc} as quasars, at lower redshifts as active galactic nuclei \cite{Merloni:2015dda}, and today in massive galaxies in their quiescent state \cite{Kormendy:2013dxa}, 
and cover a mass range from $\sim 10^4M_\odot$ \cite{2017ApJ...836..237N,2015ApJ...809L..14B,Graham:2018bre,Graham:2018byx} up to $> 10^{10}M_\odot$ \cite{McConnell:2011mu,2015Natur.518..512W,2018MNRAS.474.1342M}.
Determining the seeds of SMBHs will help us chart how they grow, and hence the role they play in the evolution of their host galaxies \cite{Ferrarese:2000se,Peng:2007mv,Volonteri:2009vh,Silk:2017yai}.
In particular, the observation of high-redshift BHs with mass $\gtrsim 100 M_\odot$, 
beyond the (pulsational) pair-instability mass gap \cite{Belczynski:2016jno,Spera:2017fyx,Woosley:2016hmi,Giacobbo:2017qhh,Marchant:2018kun}, 
would be key to understand not only the properties of very massive ($\gtrsim 250 M_\odot$) metal-poor stars \cite{Hartwig:2016nde}, 
but also the assembly of the first massive BHs in the Universe \cite{2016MNRAS.457.3356V}.

\begin{figure}[!h]
\center
    \includegraphics[width=0.491\textwidth]{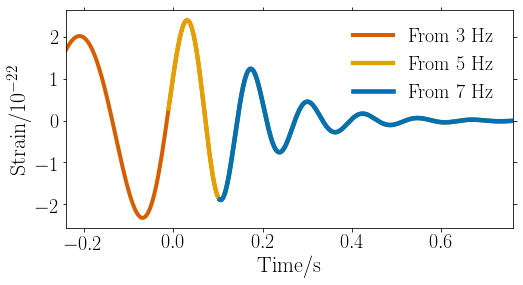}
    \includegraphics[width=0.503\textwidth]{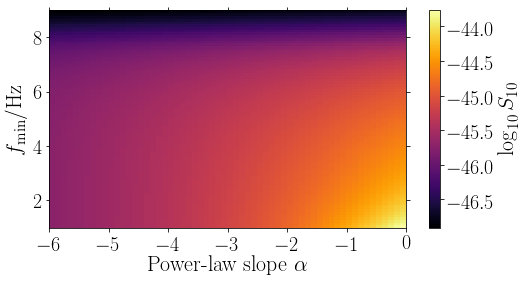}
    \caption{
\textbf{Left:}  The waveform from the final stages of inspiral, merger and ringdown of a $100 M_\odot + 100 M_{\odot}$ BBH at a redshift of $z=10$. 
Highlighted is the time evolution of the waveform from $3$, $5$ and $7~\mathrm{Hz}$. 
\textbf{Right:} Requirements on the low-frequency noise power spectrum $S_n(f)$ necessary to detect an overhead, face-on $100 M_\odot + 100 M_\odot$ BBH merging at $z=10$. 
We assume a power-law form $S_n(f) \propto f^\alpha$ extending down to a minimum frequency $f_\mathrm{min}$ with the specified normalization $S_{10}$ at $f=10~\mathrm{Hz}$.}
\label{fig:low-f}
\end{figure}

In Figure~\ref{fig:low-f}, we illustrate the importance of sensitivity in the $1$--$10~\mathrm{Hz}$ regime.  
Even with detectors sensitive to $3~\mathrm{Hz}$, we see only one cycle of a $100 M_\odot + 100 M_\odot$ circular binary with non-spinning components at $z =10$ before merger. This system is not observable above $10~\mathrm{Hz}$.  
Therefore, the objective to observe the most massive stellar-origin BBHs and the potential seeds of SMBHs early in the Universe requires new detectors sensitive to currently inaccessible frequencies below $\sim 10~\mathrm{Hz}$, which are inaccessible to current detectors.

The detectability of IMBHs places requirements on low-frequency sensitivity. 
We can model the low-frequency noise power spectral density of the detector as a power-law $S_n(f) = S_{10}(f/10~\mathrm{Hz})^\alpha$ and assume that the power law extends to some minimal frequency $f_\mathrm{min}$, below which the detectors have no sensitivity.  
In Figure~\ref{fig:low-f}, we show the combination of power law $\alpha$, minimum frequency $f_\mathrm{min}$ and the normalisation $S_{10}$ necessary to detect an optimally located and oriented merger of two $100 M_\odot$ IMBHs at $z=10$.   
There is a trade-off between the power-law slope, minimal frequency, and overall normalization, such that a range of specifications can fulfil the science requirements.  

For binaries in the currently detectable mass range, observing across a broader range of frequencies gives a more complete picture of their properties. 
The precession of component spins misaligned with the orbital angular momentum occurs over many orbits \cite{Apostolatos:1994mx,Blanchet:2013haa}. 
Its imprint is easier to discern over longer inspirals, and hence becomes more apparent with low-frequency data. 
Orbital eccentricity is rapidly damped through GW emission \cite{Peters:1964zz}.
This means that it is near unmeasurably small for current GW detectors \cite{Lower:2018seu}; however, by monitoring the earlier parts of inspiral, it will be easier to detect traces of eccentricity. 
Both the spins and the orbital eccentricity are indicative of the formation channel; enabling their measurement for large samples will have a transformative effect on our ability to answer questions about BBH origins.

\section*{Sharper -- High-precision measurements of binary properties}

Both the sensitivity and the bandwidth of next-generation detectors will enable high-precision measurements of the properties of individual binaries \cite{Vitale:2016icu,Vitale:2018nif,Hall:2019xmm}.
Parameter uncertainties are inversely proportional to the SNR \cite{Cutler:1994ys}. 
The increase in SNR made possible by the increased sensitivity will lead to exquisite measurements of the loudest events.
Increased bandwidth enables the coalescence to be tracked for a longer time, improving estimates of quantities like the spins. 
Masses, spins, merger redshifts, orbital eccentricities and (where possible) associations with host galaxies all give complementary insights into binary physics. 
High-precision measurements of individual systems allow us to make detailed studies of their origins and fundamental physics \cite{Mishra:2010tp,Gossan:2011ha,Bhagwat:2016ntk,Berti:2018vdi}. 
Combining many events together lets us study the properties of the population.  
\emph{The unique and critical advantage of GW BBH observations with next-generation detectors is the combination of high-precision measurements for a very large number of detected sources}, something that cannot be delivered by the current detectors. 

\begin{figure}[h!]
\center
  \includegraphics[width=0.98\textwidth]{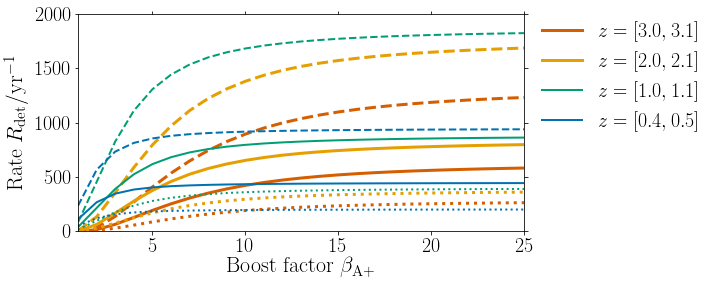}
  \caption{Expected rate of BBH detections $R_\mathrm{det}$ per redshift bin as a function of $A+$ boost factor $\beta_\mathrm{A+}$, for $z=[0.4,0.5]$, $z=[1,1.1]$, $z=[2,2.1]$, $z=[3,3.1]$.  Constant BBH merger rate densities of $53$ ($112$, $24$) $\mathrm{Gpc}^{-3}\,\mathrm{yr}^{-1}$ are shown with solid (dashed, dotted) curves, assuming equal component masses distributed according to $p(m) \propto m^{-1.6}$ \cite{LIGOScientific:2018jsj}.} 
  \label{fig:redshiftdistance}
\end{figure}

As an example, consider a highly precise reconstruction of the BH mass spectrum. 
At high masses, there is predicted to be a gap between $\simeq 45 M_{\odot}$ and $\simeq 130 M_{\odot}$ due to (pulsational) pair-instability supernova \cite{Woosley:2007qp,Woosley:2016hmi,Marchant:2018kun}. 
At lower masses, there is potentially a gap between the maximum neutron star mass and the minimum stellar BH mass \cite{Ozel:2010su,Farr:2010tu,Kreidberg:2012ud}. 
Determining the precise bounds for these gaps would provide insight into the mechanics of supernova explosions \cite{Belczynski:2011bn,Fryer:2011cx,Fryer:2017iry} and insights into the neutron star equation of state \cite{Kiziltan:2013oja,Annala:2017llu,Margalit:2017dij,Alsing:2017bbc,Abbott:2018exr}.
It can be shown \cite{Mandel:future} that: (i) for the high-mass gap, if the desired accuracy on the mass gap boundary measurement is $\sigma_g \sim 1 M_\odot$, with a conservative individual mass uncertainty for near-threshold detections of order $\sigma_m \sim 10 M_\odot$, $N \gtrsim 500$ detections are required; (ii) for the low-mass gap, $\sigma_g \sim 0.3 M_\odot$ and $\sigma_m \sim 3 M_\odot$, which would require $N \gtrsim 1500$ BBH detections. 
To provide robust answers to questions regarding massive star evolution and BBH formation, we need to trace the dependence of the boundaries of the mass gaps on metallicity and hence redshift.
Therefore, it is desirable to observe $\sim 1000$ sources in each redshift bin of width $\Delta z = 0.1$, since we may expect knowledge of the star formation rate and metallicity distribution at this resolution on the timescale of next-generation detectors \cite{Madau:2014bja}. 
Observing $1000$ sources in a given redshift bin would provide $\sim 3\%$ fractional accuracy on merger rate per redshift bin, sufficient to determine the redshift evolution of the merger rate, and constrain details of binary evolution at that redshift \cite{Zevin:2017evb,Barrett:2017fcw}.

With this in mind, we plot the number of expected BBH detections for a next-generation detector as a function of its boost factor relative to A+ in Figure~\ref{fig:redshiftdistance}. 
This assumes a BBH merger rate that does not evolve in redshift and is roughly consistent with current GW observations \cite{LIGOScientific:2018jsj}. 
From this, the target of $\sim 1000$ detections per redshift bin is achievable with boost factors of $\beta_\mathrm{A+} \sim 10$ after only $2$ years of observing time. 
These factors are possible \emph{only} with next-generation GW detectors.

\section*{Outlook for black hole gravitational-wave astronomy}

Next-generation ground-based GW detectors fulfilling the scientific objectives described here will enable the measurement of the cosmological evolution of the mass and spin distributions of BBHs and will allow us to probe their dependence on star formation history and metallicity evolution with redshift. With sensitivity increases by factors of $\sim 10$ we will be able to probe the {\em complete mass spectrum of BHs formed in merging binaries}. Such detectors will enable the robust discovery of IMBHs, if they exist, and will allow us to measure the boundaries of any mass gaps. The precise measurements of physical properties for large numbers of BH systems back to the cosmic dawn would lead to constraints on the physics of massive star evolution in single and binary systems (in connection to massive stellar winds, uncertain phases of binary interactions, as well as core-collapse supernovae and associated natal kicks), as well as to constraints on different formation channels of merging BH binaries. The potential of also revealing the nature of seed BHs for SMBHs through the unique, independent perspective of GW observations is exciting. Such data would complement those from from future electromagnetic and space-based GW observatories, enabling the maximum scientific return from these facilities.  
Next-generation GW detectors offer a unique opportunity to advance the frontiers of stellar astrophysics, the fundamental physics of compact objects, and multimessenger astronomy.

\chapterimage{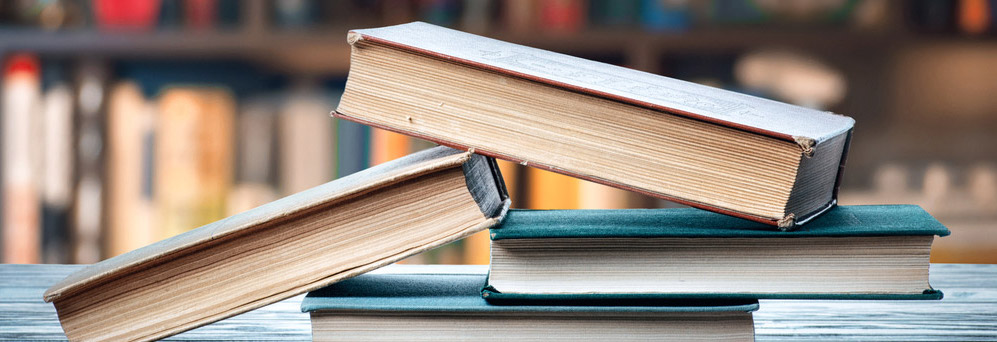} 
\bibliographystyle{utphys}
\bibliography{wp}

\end{document}